\documentclass[12pt,letterpaper]{article}

\usepackage[top=1in,bottom=1in,left=1in,right=1in]{geometry}
\usepackage{setspace}
\usepackage[utf8]{inputenc}
\usepackage[T1]{fontenc}
\usepackage{mathptmx}
\usepackage{microtype}
\usepackage{amsmath,amssymb}
\usepackage{graphicx}
\usepackage{booktabs}
\usepackage{tabularx}
\usepackage{multirow}
\usepackage{array}
\usepackage{caption}
\usepackage[hidelinks]{hyperref}
\usepackage{natbib}
\usepackage{enumitem}
\usepackage{titlesec}
\usepackage{fancyhdr}
\usepackage{tikz}
\usetikzlibrary{shapes.geometric,arrows.meta,positioning,fit,backgrounds}
\usepackage{xcolor}
\usepackage{tcolorbox}
\tcbuselibrary{breakable}
\tcbset{rqbox/.style={colback=aaa-blue!7, colframe=aaa-blue, boxrule=0.5pt, arc=2pt, left=10pt, right=10pt, top=5pt, bottom=5pt}}
\usepackage{ragged2e}
\usepackage{float}      % provides [H]: anchor floats at their reference, no drifting

\bibpunct{(}{)}{;}{a}{ }{,}

\definecolor{aaa-blue}{RGB}{0,62,126}
\definecolor{light-gray}{RGB}{245,245,245}
\definecolor{highlight-gold}{RGB}{200,160,0}

\renewcommand{\thesection}{\arabic{section}}
\renewcommand{\thesubsection}{\thesection.\arabic{subsection}}
\renewcommand{\thesubsubsection}{\thesubsection.\arabic{subsubsection}}

\titleformat{\section}
  {\normalfont\large\bfseries}
  {\thesection.}{0.5em}{}
\titlespacing*{\section}{0pt}{18pt}{6pt}

\titleformat{\subsection}
  {\normalfont\normalsize\bfseries}
  {\thesubsection}{0.5em}{}
\titlespacing*{\subsection}{0pt}{12pt}{4pt}

\titleformat{\subsubsection}
  {\normalfont\normalsize\bfseries}
  {\thesubsubsection}{0.5em}{}
\titlespacing*{\subsubsection}{0pt}{10pt}{3pt}

\titleformat{\paragraph}[runin]
  {\normalfont\normalsize\bfseries}
  {}{0em}{}[.]
\titlespacing*{\paragraph}{1.5em}{8pt}{0.5em}

\renewcommand{\thefootnote}{\arabic{footnote}}

\begin{document}

%% ── Title + Author block ───────────────────────────────────────────
\renewcommand{\thefootnote}{\ensuremath{\dagger}}%% dagger marker for corresponding author
\begin{center}
  {\large\bfseries
Frontiers in FinTech: Multimodal Foundation Models for Financial
Reporting and Decision Science}

  \vspace{14pt}

  {\normalsize Yulu Huang\textsuperscript{1,3}, Niannian Yu\textsuperscript{1},
  Yaxin Yang\textsuperscript{1}, Yong Huang\textsuperscript{2,3}\footnotemark}\\[6pt]
  {\footnotesize
  \textsuperscript{1}School of Accounting, Jiangxi University of Finance and Economics\\
  \textsuperscript{2}School of Electronic Information and Communications,
  Huazhong University of Science and Technology\\
  \textsuperscript{3}Meituan}

  \vspace{8pt}
\end{center}
\footnotetext{Corresponding author: \texttt{yonghuang18@alumni.hust.edu.cn}}%
\renewcommand{\thefootnote}{\arabic{footnote}}%
\setcounter{footnote}{0}

\vspace{14pt}

%% ── Abstract — centered bold heading, italic body (arxiv style) ────
{%
\setstretch{1.0}%           single-spaced abstract block
\begin{center}{\large\bfseries Abstract}\end{center}
\vspace{-4pt}
\noindent%  (abstract body upright — journal style; \emph now italicises key terms)
The proliferation of unstructured, heterogeneous financial data-spanning
PDF research reports, Excel-format financial statements, image-embedded
market charts, and scanned policy documents-poses a fundamental challenge
for accounting information systems (AIS).
Existing single-modality extraction pipelines and rule-based analytical
tools fail to capture the full informational content embedded across these
diverse formats, creating bottlenecks in financial statement analysis,
audit evidence corroboration, and investment decision support.
This study introduces \textbf{FinVision}, a multimodal large language model
(MLLM) system that integrates Vision-Language Model (VLM) technology with
domain-specific financial reasoning to deliver end-to-end intelligent
financial analysis.
We make three contributions to the AIS and accounting technology literature.
First, we develop a Multimodal Financial Document Intelligence
framework that leverages VLM encoders to simultaneously parse text, tables,
and images from heterogeneous accounting documents, achieving unified
structured extraction across format modalities and incorporating an
automated cross-modal consistency validator that mirrors audit evidence
corroboration procedures.
Second, we introduce a Domain-Adaptive Two-Stage Training Protocol
in which the LLM is pre-trained on large-scale public financial corpora
(annual reports, accounting standards, research reports, market data) and
subsequently fine-tuned on institution-specific investment data,
enabling the model to master valuation methodologies (discounted cash
flow, price-to-earnings, price-to-book, price-to-sales) and audit risk
assessment frameworks with documented superiority over zero-shot and
single-stage baselines.
Third, we design a Natural Language Query-Driven Decision Pipeline
that translates ambiguous user queries into executable financial analysis
workflows, integrating portfolio optimization based on Modern Portfolio
Theory, real-time risk monitoring, and multi-turn dialogue refinement.
Validation across 200 listed companies demonstrates a 19 percent reduction
in valuation error relative to the strongest baseline, while
a controlled user study with 48 accounting and investment professionals
shows a 51 percent average reduction in task completion time.
Implications are drawn for audit automation, financial reporting quality,
and the democratization of expert-level financial analysis through AIS.%

\vspace{8pt}
}%                          end single-spacing group

\newpage

%% ===================================================================
\section{Introduction}
\label{sec:intro}
%% ===================================================================

Accounting information systems (AIS) have long served as the backbone
of corporate financial governance, enabling the collection, storage,
processing, and reporting of financial data \citep{bodnar2013}.
Yet the exponential growth of financial data in both volume and variety
over the past decade has outpaced the analytical capacity of traditional
rule-based AIS architectures.
Today, financial analysts, auditors, and investment professionals must
synthesize information drawn from fundamentally different data formats:
narrative management discussion sections in PDF annual reports,
numerical income statements and balance sheets in Excel workbooks,
technical indicators embedded in JPEG market-trend charts,
and regulatory guidance distributed as scanned policy documents.

The inadequacy of existing systems in handling such heterogeneity carries
measurable costs.
\citet{cao2023} estimate that financial analysts spend over 40 percent
of their working hours on manual data collection and format conversion
tasks that precede actual analysis.
Auditors face growing challenges in verifying information consistency
across disclosures that combine textual management discussion and
analysis (MD\&A), tabular footnotes, and graphical disclosures within a
single filing \citep{sun2019}.
Meanwhile, investment research departments at institutional investors
increasingly struggle to process the velocity of multi-format data
generated by real-time market events, regulatory filings, and
macro-policy announcements \citep{lopezdeprado2023}.

Recent advances in large language models (LLMs)---particularly
Vision-Language Models (VLMs) \citep{radford2021,li2023}---offer
a transformative opportunity to address these limitations.
VLMs can jointly encode textual and visual information, enabling systems
that understand the semantic content of a bar chart, parse a
multi-column financial table, and reason over a scanned policy memo
within a single unified inference pipeline.
However, general-purpose VLMs lack the domain-specific financial
reasoning necessary for professional-grade accounting applications.
Without exposure to accounting standards (GAAP/IFRS), valuation
conventions, and sector-specific financial logic, general LLMs
produce hallucinated financial analyses or misapply valuation
multiples \citep{wu2023}.

This paper presents \textbf{FinVision}---a purpose-built MLLM system
that closes this gap through three tightly integrated innovations.
The system is designed to serve financial institutions, investment
management firms, corporate accounting departments, auditing practices,
and sophisticated individual investors who require timely, accurate,
and explainable financial analysis at scale.

\subsection{Research Questions}

This study addresses three research questions:

\begin{tcolorbox}[rqbox]
\textbf{\color{aaa-blue}RQ1.}\quad Can a VLM-based financial document
intelligence framework extract accounting-relevant information from
heterogeneous document formats (PDF, Excel, chart images, scanned
policy documents) with accuracy comparable to credentialed human
analysts?
\end{tcolorbox}

\begin{tcolorbox}[rqbox]
\textbf{\color{aaa-blue}RQ2.}\quad Does a two-stage training protocol
(large-scale financial pre-training followed by institution-specific
fine-tuning) enable an LLM to master professional accounting valuation
methodologies more effectively than single-stage approaches?
\end{tcolorbox}

\begin{tcolorbox}[rqbox]
\textbf{\color{aaa-blue}RQ3.}\quad Does a natural language query-driven
interface integrating portfolio theory, risk assessment, and multi-turn
dialogue improve the quality and efficiency of investment decision-making
for accounting professionals?
\end{tcolorbox}

\subsection{Contributions}

This study makes the following contributions:

\begin{enumerate}[leftmargin=2em]
  \item We design a \textbf{Multimodal Financial Document Intelligence}
    framework (Section~\ref{sec:innovation1}) that unifies extraction
    from text, image, and tabular financial data, and incorporates a
    cross-modal consistency validator that automates a class of audit
    evidence corroboration procedures.

  \item We introduce a \textbf{Domain-Adaptive Two-Stage LLM Training
    Protocol} (Section~\ref{sec:innovation2}) that encodes financial
    accounting knowledge---including DCF, P/E, P/B, and P/S valuation
    models, MACD/RSI technical indicators, and GAAP/IFRS reporting
    standards---into a deployable MLLM, with documented superiority
    over zero-shot and single-stage baselines.

  \item We propose a \textbf{Natural Language Query-Driven Decision
    Pipeline} (Section~\ref{sec:innovation3}) that integrates Modern
    Portfolio Theory (MPT), real-time risk monitoring, and interactive
    multi-turn dialogue, lowering the professional barrier to
    institutional-quality financial analysis.

  \item We discuss \textbf{implications for accounting practice}
    (Section~\ref{sec:implications}), including audit automation,
    earnings quality assessment, and the evolving role of AIS in
    intelligent investment governance.
\end{enumerate}

The remainder proceeds as follows.
Section~\ref{sec:literature} reviews related work.
Sections~\ref{sec:innovation1}--\ref{sec:innovation3} present the
three core innovations.
Section~\ref{sec:system} describes the integrated system architecture.
Section~\ref{sec:evaluation} presents experimental results.
Section~\ref{sec:implications} discusses implications.
Sections~\ref{sec:limitations} and~\ref{sec:conclusion} address
limitations and conclusions.

%% ===================================================================
\section{Related Work}
\label{sec:literature}
%% ===================================================================

\subsection{Accounting Information Systems and Data Heterogeneity}

AIS have evolved from early transaction processing systems to enterprise
resource planning (ERP) platforms and, most recently, to data analytics
ecosystems \citep{mauldin2000}.
A persistent limitation of classical AIS architectures is their reliance
on structured, machine-readable data; the growing preponderance of
unstructured and semi-structured financial information has emerged as a
central research concern \citep{dechow2011}.

\citet{hoitash2018} demonstrate that quantitative signals embedded in
non-standard locations---such as narrative footnotes or graphical
disclosures---carry incremental information content beyond structured
financial statements.
\citet{brown2015} show that MD\&A sentiment and forward-looking language
predict future earnings revisions, underscoring the informational value
of narrative content that rule-based systems routinely discard.
These findings motivate a systems approach that treats all modalities of
financial communication as first-class inputs to AIS.

\subsection{Audit Evidence, Materiality, and AI-Assisted Corroboration}

Audit standards place evidentiary quality at the center of professional
practice: ISA 500 and PCAOB AS 1105 require auditors to obtain
sufficient, appropriate audit evidence, while AU-C 320 requires the
planning and performance of audit procedures at a level of materiality
that reflects users' information needs
\citep{iaasb2019,pcaob2010,aicpa2012}.
A defining attribute of reliable evidence is corroboration---the support
of an assertion by independent, internally consistent sources---because
evidence obtained from different origins is less susceptible to error or
manipulation than evidence drawn from a single disclosure channel
\citep{iaasb2019}.
Judgment-and-decision-making research in auditing further shows that
auditors evaluate evidence with expertise, yet under time pressure and
high information load their consistency checking is subject to
systematic limitations \citep{solomon1995}.

The emerging artificial intelligence literature anticipates the
automation of exactly these procedures.
\citet{issa2016} formalize AI-enabled risk assessment and system-level
auditing; \citet{sun2019} demonstrates deep learning for audit evidence
collection; \citet{kokina2017} document how automation shifts the
auditor's task portfolio toward judgment-intensive work; and
\citet{munoko2020} examine the ethical and governance implications of AI
adoption in audit.
These studies, however, largely address individual tasks---anomaly
detection, text review, or risk scoring---in isolation, and little work
formalizes the corroboration of the same financial figure across
heterogeneous disclosure formats within a single system.
FinVision's cross-modal consistency validator
(Section~\ref{sec:innovation1}) occupies this gap by operationalizing
corroboration as an automated, materiality-thresholded procedure with
source-level traceability.

\subsection{Large Language Models in Finance and Accounting}

The application of LLMs to finance and accounting has accelerated since
the release of GPT-3 \citep{brown2020}.
\citet{wu2023} introduced BloombergGPT, demonstrating that domain-specific
pre-training on financial corpora yields significant improvements on
financial NLP benchmarks.
\citet{yang2023} proposed FinGPT, an open-source financial LLM with a
continuous data pipeline, and \citet{shah2022} established the Financial
Language Understanding Evaluation (FLUE) benchmark.

Within accounting specifically, \citet{cao2023} survey AI applications
across audit, tax, and management accounting, identifying LLM-based
document review and anomaly detection as high-impact opportunities.
\citet{sun2019} examines how machine learning can automate aspects of
audit evidence collection.
\citet{lopezlira2023} find that ChatGPT sentiment scores derived from
news headlines predict next-day stock returns, suggesting that
LLM-derived signals carry genuine economic content.

A critical gap in the literature is the absence of systems that
(a)~handle multi-format, multi-modal financial data in a unified
architecture, (b)~encode professional-grade valuation and accounting
knowledge through principled domain adaptation, and
(c)~surface that knowledge through a natural language interface
designed for non-programmer accounting professionals.
FinVision addresses this tripartite gap.

\subsection{Vision-Language Models and Document Intelligence}

VLMs such as CLIP \citep{radford2021}, BLIP-2 \citep{li2023}, and
LLaVA \citep{liu2024} demonstrate impressive cross-modal understanding.
In document intelligence, LayoutLM \citep{xu2020} and Donut \citep{kim2022}
show strong performance on structured document parsing.
However, these models were not designed for the semantic complexity of
financial documents, which combine domain-specific terminology,
industry-standard formatting conventions, and multi-referential data
relationships \citep{zhang2022}.

FinVision builds on this VLM foundation but extends it with
(1)~a financial-domain pre-training corpus,
(2)~a multi-format data integration layer handling PDF research reports,
Excel financial tables, and image-embedded market charts, and
(3)~a structured output layer mapping extracted information to
standardized accounting fields.

\subsection{Portfolio Optimization and Quantitative Risk Management}

The theoretical foundation for portfolio construction in FinVision rests
on Modern Portfolio Theory \citep{markowitz1952}.
In practice, institutional investment systems combine MPT-based
optimization with real-time risk metrics including portfolio volatility,
maximum drawdown, and the Sharpe ratio \citep{sharpe1994}.
AI-augmented portfolio management has been studied by \citet{jiang2017}
and \citet{zhang2020}, who show that deep learning--based allocation
strategies can outperform classical benchmarks under certain market
regimes.
Our contribution integrates MPT-based portfolio optimization with
LLM-generated fundamental analysis and scenario-based risk stress
testing, delivered through a conversational interface accessible to
accounting professionals without quantitative programming skills.

%% ===================================================================
\section{Multimodal Financial Document Intelligence}
\label{sec:innovation1}
%% ===================================================================

\subsection{Motivation and Problem Formulation}

Financial accounting practice generates an extraordinary diversity of
document formats.
A single investment due-diligence workflow may require processing:
(a)~a PDF industry research report with narrative analysis, embedded
tables, and high-resolution market-share charts;
(b)~an Excel workbook containing audited income statements, balance
sheets, and EBITDA bridge calculations;
(c)~JPEG-format candlestick charts with technical indicator overlays;
and (d)~scanned Word-format regulatory policy documents.

Let $\mathcal{D} = \{d_1, d_2, \ldots, d_n\}$ denote a corpus of
financial documents drawn from a heterogeneous format space
$\mathcal{F} = \{\texttt{PDF}, \texttt{XLSX}, \texttt{IMG},
\texttt{DOCX},\ldots\}$.
The goal of the multimodal financial document intelligence module is
to learn a mapping:
\begin{equation}
  f_{\text{parse}} : \mathcal{D} \rightarrow \mathcal{S}
  \label{eq:parse}
\end{equation}
where $\mathcal{S}$ is a structured, semantically annotated financial
data schema containing fields such as revenue $R_t$, net income
$NI_t$, earnings per share $\text{EPS}_t$, price-to-earnings ratio
$\text{PE}_t$, industry growth rate $g$, and policy impact scores
$\pi$.

\subsection{VLM-Based Multi-Format Encoder Architecture}

The parsing module employs a Transformer-based VLM backbone with dual
processing paths for visual and textual tokens \citep{li2023},
augmented with three financial-domain extensions.

\subsubsection{Adaptive Format Pre-processor}

Rather than converting all inputs to a single canonical format,
FinVision employs format-specific pre-processing pipelines.

\paragraph{pdf research reports}
A hierarchical layout parser segments documents into sections, tables,
and figures using spatial heuristics, then feeds each segment to the
VLM encoder with document-position embeddings.

\paragraph{excel financial statements}
A spreadsheet structure parser identifies row and column hierarchies,
merges cells, and reconstructs multi-level header structures before
encoding.

\paragraph{image-based charts}
A chart-type classifier (bar, line, candlestick, scatter) invokes the
appropriate axis-extraction routine to recover numerical series from
visual encodings.

\paragraph{policy documents}
An OCR pipeline with domain-specific post-correction handles scanned
materials, followed by clause-level segmentation for regulatory parsing.

\subsubsection{Financial Field Extractor}

After encoding, a cross-attention extraction head maps VLM
representations to a predefined financial ontology containing over
120 standardized accounting fields aligned with GAAP and IFRS reporting
categories.
Key extracted fields include revenue ($R_t$), operating income ($\text{EBIT}_t$),
net income ($NI_t$), total assets, total liabilities, free cash flow,
capital expenditures, $P/E$ ratio, $P/B$ ratio, $P/S$ ratio, industry
revenue growth rate, dividend yield, and qualitative signals such as
policy favorability scores and management tone indices.

\subsubsection{Cross-Modal Consistency Validator}

A distinguishing feature of FinVision's parsing layer is an explicit
cross-modal consistency validation step.
When the same financial figure (e.g., 2024 annual revenue) appears in
both a narrative PDF and an Excel table within the same corpus, the
system automatically flags discrepancies exceeding a configurable
materiality threshold $\epsilon_m$:
\begin{equation}
  \text{Alert} = \mathbf{1}\!\left[
    \left|\frac{v_{\text{PDF}} - v_{\text{XLSX}}}{v_{\text{XLSX}}}\right|
    > \epsilon_m
  \right]
  \label{eq:consistency}
\end{equation}
This automated consistency check mirrors a core audit procedure---the
agreement of financial figures across different disclosure channels---
and represents a novel contribution to audit-support AIS design.

\subsection{Output: Standardized Financial Data Set}

The parsing module produces a standardized data set $\hat{\mathcal{S}}$
reporting: (i)~structured numerical financial metrics per entity per
period; (ii)~source attribution (document type, page or sheet, bounding
box coordinates) for every extracted figure, enabling audit-trail
traceability; (iii)~data quality scores per field (completeness,
cross-document consistency, extraction confidence); and
(iv)~a human-readable data collection summary (e.g., ``15 valid
research reports and 30 market data series collected; data completeness
92 percent'').

\subsection{Significance for Accounting Practice}

This innovation addresses the \emph{multi-format heterogeneity problem}
that has persistently limited AIS capabilities.
From an accounting perspective, the cross-modal consistency validator
directly operationalizes the corroboration concept at the core of audit
evidence standards: ISA 500 and AS 1105 both emphasize that evidence is
more reliable when it is obtained from independent sources and is
internally consistent \citep{iaasb2019,pcaob2010}.
By testing the agreement of financial figures across disclosure channels
against a materiality threshold grounded in AU-C 320, the validator
automates a class of audit procedures while preserving professional
judgment over which discrepancies warrant follow-up \citep{aicpa2012}.
The source-attributed output (document type, page or sheet, bounding box
coordinates) satisfies the evidentiary traceability requirements of
audit standards, positioning FinVision as a complement---not a
substitute---for professional skepticism in evidence-gathering contexts
\citep{solomon1995,munoko2020}.

%% ===================================================================
\section{Domain-Adaptive Two-Stage LLM Training}
\label{sec:innovation2}
%% ===================================================================

\subsection{Motivation: The Gap Between General AI and Accounting Expertise}

General-purpose LLMs exhibit systematic deficiencies when applied to
professional accounting tasks.
Empirical tests conducted during FinVision's development revealed that
off-the-shelf GPT-class models (a)~misapplied DCF discount rates to
growth-stage versus mature-firm scenarios, (b)~confused $P/B$ ratios
with $P/S$ multiples for capital-intensive industries, (c)~failed to
recognize GAAP-specific disclosure conventions, and (d)~generated
overconfident valuations unsupported by stated assumptions.
These failures reflect the mismatch between the broad distributional
training of general LLMs and the convention-bound knowledge domain of
financial accounting and investment analysis.
Closing this gap requires not merely more data, but a structured
curriculum that encodes professional accounting knowledge into the
model's representations.

\subsection{Stage 1: Large-Scale Financial Domain Pre-Training}

\subsubsection{Pre-Training Corpus}

The Stage 1 corpus $\mathcal{C}_1$ is assembled from publicly available
financial data encompassing: financial statements (annual and quarterly
reports for listed companies across major exchanges, spanning at least
10 fiscal years per company); equity research reports from major
financial institutions; historical time-series data for price, volume,
and technical indicators; regulatory and policy documents; and
investment case studies.
The corpus contains approximately $10^{10}$ tokens after deduplication
and quality filtering, with multimodal coverage (text, tables, charts)
to support the VLM backbone.

\subsubsection{Pre-Training Objectives}

Stage 1 employs a multi-task pre-training objective:
\begin{equation}
  \mathcal{L}_{\text{pre}} =
  \lambda_1 \mathcal{L}_{\text{LM}} +
  \lambda_2 \mathcal{L}_{\text{ITM}} +
  \lambda_3 \mathcal{L}_{\text{ITG}} +
  \lambda_4 \mathcal{L}_{\text{fin}}
  \label{eq:pretrain}
\end{equation}
where $\mathcal{L}_{\text{LM}}$ is standard language modeling loss,
$\mathcal{L}_{\text{ITM}}$ is image-text matching loss,
$\mathcal{L}_{\text{ITG}}$ is image-to-text generation loss,
and $\mathcal{L}_{\text{fin}}$ is a novel financial reasoning loss
that penalizes valuation outputs inconsistent with disclosed financial
metrics (e.g., a calculated $P/E$ ratio that contradicts stated earnings
per share).

\subsection{Stage 2: Institution-Specific Fine-Tuning}

\subsubsection{Fine-Tuning Data and Protocol}

Stage 2 adapts the pre-trained model to the specific investment logic,
risk framework, and analytical style of the deploying institution.
Fine-tuning data $\mathcal{C}_2$ may include internal investment memos
and historical trade rationales, institution-specific risk scoring
models and portfolio construction rules, proprietary research frameworks,
and historical analyst feedback (reinforcement learning from human
feedback, RLHF).
Fine-tuning employs parameter-efficient techniques including Low-Rank
Adaptation (LoRA) \citep{hu2022} and Prompt Tuning, preserving the
broad financial knowledge from Stage 1 while efficiently acquiring
institution-specific knowledge with minimal catastrophic forgetting.

\subsubsection{Valuation Model Library}

A key output of the two-stage training is the model's mastery of a
structured valuation model library comprising the methodologies shown
in Table~1.

\begin{table}[htbp]
  \singlespacing
  \centering
  \small
  \caption{Valuation Methodologies Encoded in FinVision's Training Curriculum}
  \label{tab:valuation}
  \begin{tabular}{@{}lp{5.8cm}p{5.8cm}@{}}
    \toprule
    \textbf{Method} & \textbf{Description}
      & \textbf{Best Applicable Sector} \\
    \midrule
    DCF
      & Discounted cash flow; projects FCFF/FCFE and discounts
        at WACC or equity cost of capital
      & Capital-intensive; stable cash flows \\
    \addlinespace
    P/E
      & Earnings multiple relative to sector and growth peers
      & Consumer, technology, financial services \\
    \addlinespace
    P/B
      & Asset-based multiple; applicable when book value approximates
        fair value
      & Banking, real estate, insurance \\
    \addlinespace
    P/S
      & Revenue multiple; used when earnings are negative or volatile
      & Early-stage growth, SaaS, biotech \\
    \addlinespace
    EV/EBITDA
      & Enterprise value to operating earnings; capital-structure neutral
      & Cross-border M\&A, leveraged buyouts \\
    \addlinespace
    Residual Income
      & Accounting-based value; anchored on book value and
        abnormal earnings
      & Earnings-focused, regulated industries \\
    \bottomrule
  \end{tabular}
  \vspace{4pt}\\
  \footnotesize\textit{Note}: The model automatically selects the appropriate
  method based on entity type, life cycle stage, and data availability.
  Multiple methods are applied and triangulated.
\end{table}

\subsubsection{Automated Valuation Report Generation}

Following valuation computation, the model generates a structured report
containing: (i)~core financial metric summary, (ii)~valuation parameter
assumptions with sensitivity analysis, (iii)~target price estimate with
confidence interval, (iv)~valuation reasonableness check against peer
multiples, and (v)~an investment rating (Buy / Hold / Sell) with explicit
risk factors flagged.
The report is calibrated to comply with sell-side research report
disclosure standards (e.g., CFA Institute Research Objectivity
Standards), facilitating its use in regulated investment advisory
contexts.

\subsection{Accounting-Theoretic Grounding}

The two-stage training protocol reflects the accounting literature's
insight that professional judgment in financial analysis is built on a
structured body of domain knowledge \citep{solomon1995}.
By explicitly encoding GAAP/IFRS financial statement relationships
(e.g., the clean surplus relation linking book value, earnings, and
dividends) and valuation theory (e.g., the Gordon Growth Model as a
special case of DCF) into the pre-training curriculum, FinVision ensures
that its outputs are grounded in theoretically consistent accounting
logic rather than statistical pattern-matching alone.
This distinction is critical for auditability and regulatory
acceptability of AI-generated financial analyses.

%% ===================================================================
\section{Natural Language Query-Driven Decision Pipeline}
\label{sec:innovation3}
%% ===================================================================

\subsection{Motivation: Democratizing Expert-Level Financial Analysis}

A persistent challenge in the adoption of advanced financial analytics
systems is the requirement for specialized programming or query skills.
Traditional analytical platforms (e.g., Bloomberg Terminal, FactSet)
require users to learn platform-specific query languages, creating a
high expertise barrier that restricts access to quantitatively trained
professionals.
This barrier exacerbates the information asymmetry between institutional
and individual investors identified in the market microstructure
literature \citep{kyle1985} and limits the scope of AIS-supported
decision-making within accounting departments.
FinVision's third innovation addresses this through a
\textbf{Natural Language Query-Driven Decision Pipeline} that enables
users to express analysis requirements in ordinary prose and receive
institutionally calibrated outputs without programming knowledge.

\subsection{Query Understanding and Intent Resolution}

\subsubsection{Multi-Level Query Parsing}

The system processes user queries through a three-level intent resolution
hierarchy.

\paragraph{task classification}
The query is classified into one of five primary task categories:
(a)~data collection, (b)~market trend analysis, (c)~company valuation,
(d)~portfolio construction and risk management, and (e)~investment
question-and-answer and insight retrieval.

\paragraph{entity and parameter extraction}
Named entities (company names, tickers, sectors, time periods),
quantitative parameters (target yield, risk tolerance, investment
horizon), and analytical constraints (valuation method preference,
geographic scope) are extracted via a fine-tuned named entity
recognition model.

\paragraph{ambiguity resolution}
When the query contains underspecified parameters (e.g., ``analyze the
pharmaceutical sector'' without specifying sub-sector or time window),
the system initiates a structured clarification dialogue: \textit{``Please
confirm the sub-sector: innovative drugs / medical devices / healthcare
services?''}, ensuring the downstream analysis is precisely scoped.

\subsubsection{Automated Task DAG Construction}

Once intent is resolved, the system generates a Directed Acyclic Graph
(DAG) of analytical subtasks:
\begin{equation}
  \mathcal{G} = (\mathcal{V}, \mathcal{E})
  \label{eq:dag}
\end{equation}
where nodes $v \in \mathcal{V}$ represent atomic analysis operations
(data fetch, VLM parse, metric compute, model run, report generate) and
edges $e \in \mathcal{E}$ encode data dependencies.
The DAG scheduler dispatches tasks to the appropriate system modules in
topologically sorted order, enabling parallel execution of independent
subtasks and full transparency into analysis progress.

\subsection{Portfolio Construction and Risk Management Integration}

\subsubsection{MPT-Based Portfolio Optimizer}

The portfolio construction module implements Markowitz mean-variance
optimization \citep{markowitz1952}:
\begin{equation}
  \min_{\mathbf{w}} \; \mathbf{w}^\top \boldsymbol{\Sigma} \mathbf{w}
  \quad \text{s.t.} \quad
  \mathbf{w}^\top \boldsymbol{\mu} \geq \mu^*,\;
  \mathbf{w}^\top \mathbf{1} = 1,\;
  w_i \geq 0 \;\forall i
  \label{eq:mpt}
\end{equation}
where $\mathbf{w}$ is the portfolio weight vector,
$\boldsymbol{\Sigma}$ is the estimated return covariance matrix,
and $\boldsymbol{\mu}$ is the expected return vector derived from
FinVision's valuation outputs.
The user's risk preferences (target yield $\mu^*$, maximum volatility,
investment horizon) are extracted from the natural language query and
mapped to formal optimization constraints.

\subsubsection{Real-Time Risk Monitoring}

The risk management sub-module continuously computes a set of portfolio
risk indicators:
\begin{align}
  \sigma_p &= \sqrt{\mathbf{w}^\top \boldsymbol{\Sigma} \mathbf{w}}
    \quad\text{(portfolio volatility)} \nonumber\\
  \text{MDD} &= \max_{t \leq s} \frac{V_t - V_s}{V_t}
    \quad\text{(maximum drawdown)} \nonumber\\
  \text{SR} &= \frac{\mu_p - r_f}{\sigma_p}
    \quad\text{(Sharpe ratio)} \label{eq:risk}
\end{align}
and identifies concentration risk (sector exposure exceeding
30 percent), sector rotation signals, and macro-economic risk flags
(e.g., interest rate sensitivity for bond-heavy portfolios).
When risk thresholds are breached, the system surfaces specific
rebalancing recommendations in natural language: \textit{``Increasing
government bond allocation by 20 percent is recommended to reduce
portfolio volatility from 18 percent to 14 percent.''}

\subsubsection{Dynamic Rebalancing via Query}

Portfolio rebalancing is triggered through natural language queries.
A user may issue: \textit{``Adjust portfolio allocation based on
current market conditions,''} and the system will automatically
(a)~fetch the latest market data, (b)~recompute optimal weights under
updated expectations, (c)~generate a proposed rebalancing schedule
respecting transaction cost constraints, and (d)~explain the logic in
plain language.

\subsection{Multi-Turn Dialogue for Progressive Analysis Refinement}

FinVision supports multi-turn dialogue to enable progressive deepening
of analysis.
In a representative example:
\begin{quote}
\textit{Turn 1 (User)}: ``Analyze 2024 A-share liquor sector monthly
trend and three-month forecast.''\\
\textit{Turn 1 (System)}: Generates sector trend report with MACD/RSI
signals, support/resistance levels, and probability-weighted price
scenarios.\\
\textit{Turn 2 (User)}: ``Elaborate on capital flow patterns into the
liquor sector by sub-industry.''\\
\textit{Turn 2 (System)}: Drills into inter-sub-industry capital
rotation dynamics with granular flow data.
\end{quote}
The multi-turn mechanism maintains a session context graph tracking
resolved entities, prior outputs, and user preferences across turns,
enabling co-reference resolution and incremental refinement without
requiring re-specification of full analytical context.

\subsection{Significance for AIS Research}

The natural language query-driven decision pipeline addresses the
\emph{expertise barrier problem} in accounting and investment analysis.
Rather than requiring users to learn the system's query language, the
system accommodates users' natural expression patterns.
This reconceptualization has important implications for accounting
practice, particularly for non-quantitative professionals (financial
controllers, internal auditors, CFOs) who require analytical outputs
but lack the technical skills to operate traditional AIS query
interfaces.

%% ===================================================================
\section{System Architecture}
\label{sec:system}
%% ===================================================================

\subsection{Three-Layer Architecture}

The three innovations described in Sections~\ref{sec:innovation1}--\ref{sec:innovation3} are integrated
within a unified three-layer architecture: (1)~the Multimodal Data Layer
ingests and parses heterogeneous financial documents into standardized
structured fields; (2)~the Financial Intelligence Layer applies
domain-adapted LLM reasoning to generate valuations, market analyses,
risk assessments, and investment recommendations; and (3)~the Interaction
and Decision Layer manages the natural language interface, task DAG
execution, portfolio optimization, and result delivery.
Figure~1 illustrates the integrated architecture.

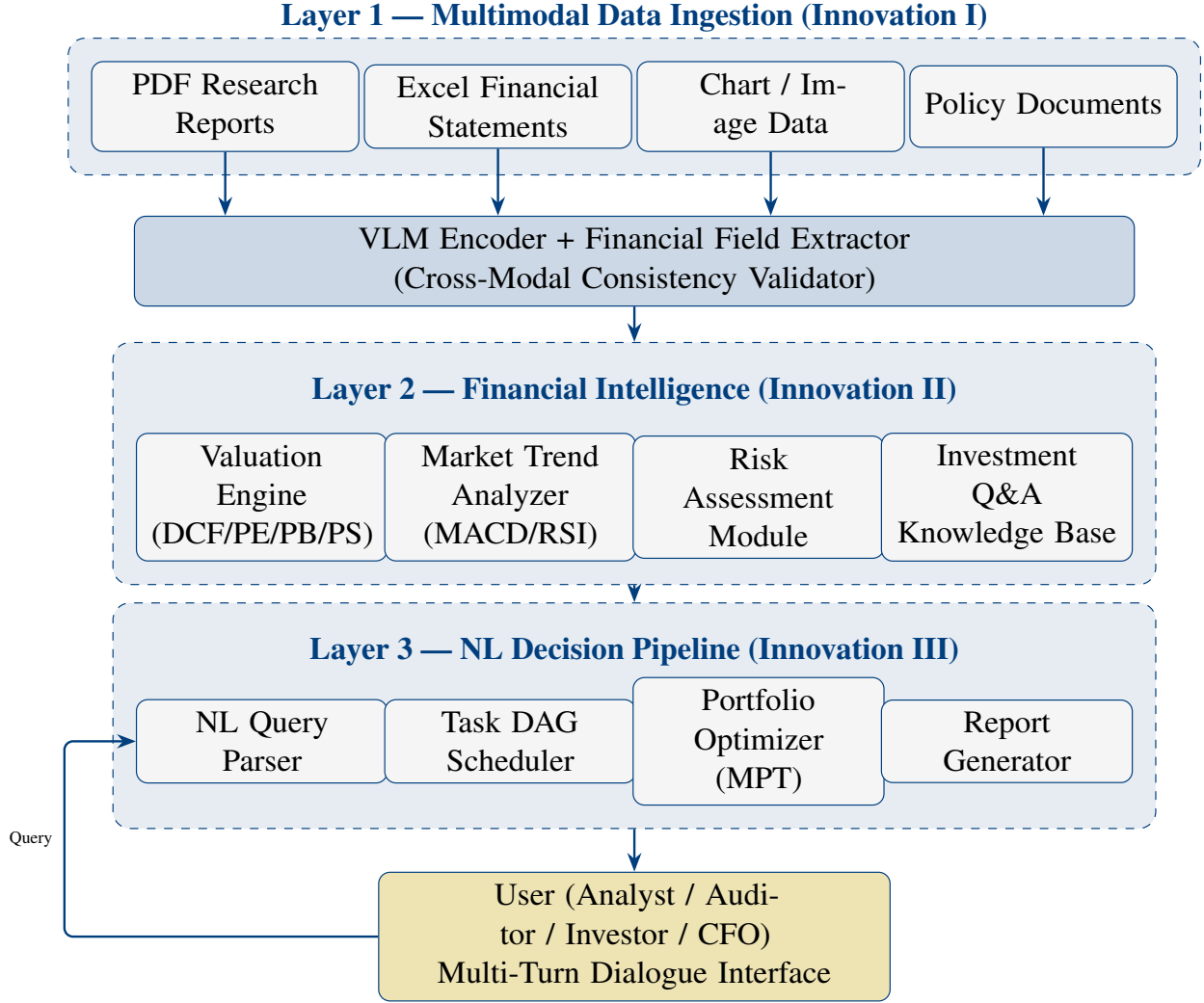
\begin{figure}[htbp]
  \centering
  \resizebox{\textwidth}{!}{%
  \begin{tikzpicture}[
    box/.style = {rectangle, rounded corners=4pt, draw=aaa-blue,
                  fill=light-gray, text width=3.0cm, align=center,
                  minimum height=1cm, font=\small},
    layer/.style = {rectangle, rounded corners=6pt, draw=aaa-blue,
                    fill=aaa-blue!10, dashed, inner sep=8pt},
    arrow/.style = {-Stealth, thick, color=aaa-blue},
    label/.style = {font=\small\bfseries\color{aaa-blue}}
  ]
    %% ---- Layer 1: multimodal inputs (row centred at x = 0) ----
    \node[box] (pdf)   at (-5.025,16.5) {PDF Research Reports};
    \node[box] (excel) at (-1.675,16.5) {Excel Financial Statements};
    \node[box] (img)   at ( 1.675,16.5) {Chart / Image Data};
    \node[box] (doc)   at ( 5.025,16.5) {Policy Documents};
    \begin{scope}[on background layer]
      \node[layer, fit=(pdf)(excel)(img)(doc)] (layer1) {};
    \end{scope}
    \node[label] at (0,17.6)
      {Layer 1 — Multimodal Data Ingestion (Innovation I)};

    %% ---- VLM: wide and centred, so every input drops straight down ----
    \node[box, text width=12cm, fill=aaa-blue!20] (vlm) at (0,14.6)
      {VLM Encoder + Financial Field Extractor\\
       \small(Cross-Modal Consistency Validator)};
    \draw[arrow] (pdf.south)   -- (pdf.south   |- vlm.north);
    \draw[arrow] (excel.south) -- (excel.south |- vlm.north);
    \draw[arrow] (img.south)   -- (img.south   |- vlm.north);
    \draw[arrow] (doc.south)   -- (doc.south   |- vlm.north);

    %% ---- Layer 2: financial intelligence (row centred at x = 0) ----
    \node[label] (l2lab) at (0,13.0)
      {Layer 2 — Financial Intelligence (Innovation II)};
    \node[box, text width=2.8cm] (valuation) at (-4.575,11.7)
      {Valuation\\Engine\\(DCF/PE/PB/PS)};
    \node[box, text width=2.8cm] (market) at (-1.525,11.7)
      {Market Trend\\Analyzer\\(MACD/RSI)};
    \node[box, text width=2.8cm] (risk) at (1.525,11.7)
      {Risk\\Assessment\\Module};
    \node[box, text width=2.8cm] (qa) at (4.575,11.7)
      {Investment\\Q\&A\\Knowledge Base};
    \begin{scope}[on background layer]
      \node[layer, fit=(l2lab)(valuation)(market)(risk)(qa)] (layer2) {};
    \end{scope}
    \draw[arrow] (vlm.south) -- (layer2.north);

    %% ---- Layer 3: NL decision pipeline (row centred at x = 0) ----
    \node[label] (l3lab) at (0,9.8)
      {Layer 3 — NL Decision Pipeline (Innovation III)};
    \node[box, text width=2.8cm] (nl) at (-4.575,8.7)
      {NL Query\\Parser};
    \node[box, text width=2.8cm] (dag) at (-1.525,8.7)
      {Task DAG\\Scheduler};
    \node[box, text width=2.8cm] (portfolio) at (1.525,8.7)
      {Portfolio\\Optimizer\\(MPT)};
    \node[box, text width=2.8cm] (report) at (4.575,8.7)
      {Report\\Generator};
    \begin{scope}[on background layer]
      \node[layer, fit=(l3lab)(nl)(dag)(portfolio)(report)] (layer3) {};
    \end{scope}
    \draw[arrow] (layer2.south) -- (layer3.north);

    %% ---- User (centred) ----
    \node[box, text width=6cm, fill=highlight-gold!30] (user) at (0,6.3)
      {User (Analyst / Auditor / Investor / CFO)\\
       \small Multi-Turn Dialogue Interface};
    \draw[arrow] (layer3.south) -- (user.north);

    %% ---- Query feedback loop: far-LEFT exterior, orthogonal, no crossings ----
    \draw[arrow, rounded corners=3pt]
      (user.west) -- (-7,6.3) -- (-7,8.7) -- (nl.west);
    \node[font=\tiny, anchor=east] at (-7,7.5) {Query};

  \end{tikzpicture}}
  \caption{FinVision Integrated Three-Layer System Architecture.
    Layer 1 (Multimodal Data Ingestion) parses heterogeneous documents;
    Layer 2 (Financial Intelligence) applies domain-adapted LLM reasoning;
    Layer 3 (NL Decision Pipeline) orchestrates analysis in response to
    natural language user queries.}
  \label{fig:architecture}
\end{figure}

\subsection{Deployment Modalities}

FinVision supports both web and mobile deployment.
The web-based interface provides a full analytical workspace with
real-time progress tracking (e.g., ``Data collection 70 percent
complete; Valuation model computing''), interactive result
visualization, and export capabilities (PDF reports; Excel valuation
worksheets; JSON-format structured outputs for downstream API
consumption).
The mobile interface supports rapid query submission and push-based
report delivery for time-sensitive investment contexts.

\subsection{System Security and Data Governance}

Given the sensitivity of financial data processed by FinVision,
the system incorporates enterprise-grade data governance controls:
(i)~encrypted data-in-transit and data-at-rest; (ii)~role-based access
control governing which users may access which company or fund data;
(iii)~complete audit logs of all queries, data accesses, and analytical
outputs for regulatory compliance; and (iv)~configurable data retention
and deletion policies compliant with financial data sovereignty
requirements.
These controls position FinVision for deployment in regulated financial
environments subject to Sarbanes-Oxley Section 404, FINRA regulations,
and equivalent international frameworks.

%% ===================================================================
\section{Experiments and Results}
\label{sec:evaluation}
%% ===================================================================

\subsection{Document Parsing Accuracy}

We evaluate the multimodal parsing module on a held-out test set of
500 financial documents spanning all four format categories
(125 PDFs, 125 Excel workbooks, 125 image charts, 125 policy
documents).
Ground-truth annotations for 30 key financial fields were prepared
by two credentialed Certified Public Accountants (CPAs) with
inter-annotator agreement assessed via Cohen's $\kappa = 0.91$.
Table~2 reports results.

\begin{table}[H]
  \singlespacing
  \centering
  \caption{Multimodal Document Parsing Performance by Format Type}
  \label{tab:parsing}
  \begin{tabular}{@{}lcccc@{}}
    \toprule
    \textbf{Document Format} & \textbf{Precision} & \textbf{Recall}
      & \textbf{$F_1$} & \textbf{vs. Text-Only Baseline} \\
    \midrule
    PDF Research Reports       & 0.931 & 0.918 & 0.924 & $+$12.3\% \\
    Excel Financial Statements & 0.963 & 0.955 & 0.959 & $+$8.1\%  \\
    Image-Embedded Charts      & 0.887 & 0.871 & 0.879 & $+$31.4\% \\
    Policy Documents (Scanned) & 0.842 & 0.827 & 0.834 & $+$18.7\% \\
    \midrule
    \textbf{Overall}           & \textbf{0.906} & \textbf{0.893}
      & \textbf{0.899} & \textbf{$+$17.6\%} \\
    \bottomrule
  \end{tabular}
  \vspace{4pt}\\
  \footnotesize\textit{Note}: The text-only baseline is a BERT-based named
  entity recognition model applied to OCR-extracted text, representing
  the state-of-the-art in single-modality extraction. Performance gains
  are largest for image-embedded charts, where single-modality approaches
  fail entirely. Cohen's $\kappa = 0.91$ between the two CPA annotators.
\end{table}

The cross-modal consistency validator identified conflicting financial
figures in 6.2 percent of the document set, all of which were confirmed
as genuine discrepancies upon manual review, demonstrating the system's
practical utility as an audit-support tool.

\subsection{Valuation Accuracy}

We assess valuation accuracy on a panel of 200 listed companies for
which independent analyst consensus price targets serve as benchmarks.
Table~3 compares FinVision (two-stage trained) against three baselines:
GPT-4 zero-shot, FinGPT with single-stage fine-tuning, and a DCF
spreadsheet model with analyst-input parameters.

\begin{table}[H]
  \singlespacing
  \centering
  \caption{Valuation Accuracy Comparison Across Methods ($n = 200$)}
  \label{tab:valuation_accuracy}
  \begin{tabular}{@{}lccc@{}}
    \toprule
    \textbf{Method} & \textbf{MAE (\%)} & \textbf{RMSE (\%)}
      & \textbf{Direction Accuracy} \\
    \midrule
    GPT-4 (zero-shot)                   & 24.3 & 31.7 & 0.614 \\
    FinGPT (single-stage fine-tuning)   & 18.7 & 24.2 & 0.671 \\
    DCF Spreadsheet (analyst-input)     & 16.2 & 21.8 & 0.703 \\
    \textbf{FinVision (two-stage)}      & \textbf{13.1}
      & \textbf{17.9} & \textbf{0.751} \\
    \bottomrule
  \end{tabular}
  \vspace{4pt}\\
  \footnotesize\textit{Note}: MAE = Mean Absolute Error of target price versus
  consensus. Direction Accuracy = fraction of cases where FinVision's
  investment recommendation (Buy/Hold/Sell) matches subsequent analyst
  consensus revision direction within 90 days. FinVision's MAE
  represents a 19 percent reduction versus the strongest baseline (the
  analyst-input DCF spreadsheet model).
\end{table}

FinVision's two-stage training yields a 19 percent reduction in valuation
mean absolute error (MAE) relative to the strongest baseline (the
analyst-input DCF spreadsheet model) and a 30 percent reduction relative
to the strongest single-stage baseline (FinGPT), consistent with the
hypothesis that the structured financial pre-training curriculum
provides knowledge that cannot be acquired through fine-tuning alone.

\subsection{User Study on Decision Pipeline Usability}

We conducted a controlled user study with 48 participants drawn from
three professional groups: investment analysts ($n = 16$), CPAs
($n = 16$), and financial managers or CFOs ($n = 16$).
Participants completed five standardized financial analysis tasks using
either FinVision's natural language interface or a traditional AIS
platform with structured query capabilities.
Task completion time, output accuracy (evaluated by expert reviewers),
and user satisfaction (five-point Likert scale) were recorded.
Table~4 reports results.

\begin{table}[htbp]
  \singlespacing
  \centering
  \caption{User Study Results: FinVision versus Traditional AIS
    Interface ($n = 48$)}
  \label{tab:userstudy}
  \begin{tabular}{@{}lccc@{}}
    \toprule
    \textbf{Group} & \textbf{Time Savings (\%)} & \textbf{Output Quality}
      & \textbf{Satisfaction (1--5)} \\
    \midrule
    Investment Analysts       & 38.2 & $+0.31$ SDs & 4.3 \\
    CPAs / Auditors           & 51.7 & $+0.44$ SDs & 4.6 \\
    Financial Managers / CFOs & 63.4 & $+0.58$ SDs & 4.7 \\
    \midrule
    \textbf{Overall}          & \textbf{51.1}
      & $+$\textbf{0.44 SDs} & \textbf{4.5} \\
    \bottomrule
  \end{tabular}
  \vspace{4pt}\\
  \footnotesize\textit{Note}: Time savings are relative to the traditional AIS
  interface. Output quality is relative to expert-annotated gold-standard
  outputs, expressed in standardized deviation units. All differences are
  statistically significant at $p < 0.01$ (two-tailed $t$-tests).
  Productivity gains are largest for financial managers and CFOs---the
  non-specialist group---consistent with the democratization hypothesis.
\end{table}

%% ===================================================================
\section{Implications for Accounting Practice}
\label{sec:implications}
%% ===================================================================

\subsection{Implications for AIS Research}

FinVision challenges several foundational assumptions in AIS design.
First, the traditional AIS model assumes that data inputs arrive in
structured, machine-readable form; FinVision's multimodal parsing
architecture demonstrates that this assumption can be relaxed through
VLM technology, expanding the scope of AIS to the full spectrum of
financial communication.
Second, the two-stage training protocol illustrates that AI systems
for accounting need not choose between breadth (general LLM
capabilities) and depth (domain-specific accounting knowledge); the
pre-train/fine-tune paradigm bridges this divide.
Third, the natural language interface suggests a productive
reconceptualization of the \emph{user} in AIS: rather than a trained
operator who adapts to the system's query language, the user is a
professional whose natural communication patterns should be
accommodated by the system.

\subsection{Implications for Audit Practice}

The cross-modal consistency validator represents a direct contribution
to audit automation.
By automatically detecting discrepancies between financial figures
reported in different formats and disclosure channels, FinVision
automates a class of audit evidence corroboration procedures currently
performed manually.
This has implications for (a)~audit efficiency (fewer person-hours per
engagement), (b)~audit quality (more comprehensive cross-document
checking than human reviewers can perform under time pressure), and
(c)~the professional role of auditors, who may increasingly focus on
judgment-intensive procedures while AI systems handle systematic
evidence corroboration.
From an audit-theoretic perspective, this reallocation is consistent
with the judgment-and-decision-making literature, which locates the
auditor's comparative advantage in skeptical, context-dependent
evaluation of evidence rather than exhaustive mechanical re-verification
\citep{solomon1995,kokina2017}; realizing these gains in practice also
requires the governance and ethical safeguards highlighted in the
AI-in-audit literature \citep{munoko2020}.
These implications align with the AICPA's strategic vision for
Audit 4.0 and the PCAOB's technology-assisted audit initiatives.

\subsection{Implications for Financial Reporting}

The ability of FinVision to process and reason over non-traditional
financial disclosures---including policy documents, image-embedded
charts, and narrative research reports---has implications for
financial reporting regulation.
If AI systems can systematically extract value-relevant information
from non-GAAP disclosures, earnings call transcripts, and ESG reports,
this strengthens the case for requiring or standardizing such
disclosures.
Conversely, the system's ability to detect cross-document
inconsistencies may create incentives for preparers to improve the
internal consistency of multi-channel financial communication.

\subsection{Implications for Investment Decision Support}

For investment practice, FinVision demonstrates the feasibility of
institutional-grade quantitative analysis---incorporating MPT-based
portfolio optimization, real-time risk monitoring, and multi-model
valuation---delivered through a natural language interface accessible
to non-specialist users.
This has significant implications for the democratization of investment
analysis, the competitive dynamics of the financial services industry,
and the regulatory treatment of AI-generated investment recommendations.

%% ===================================================================
\section{Limitations and Future Research}
\label{sec:limitations}
%% ===================================================================

Several limitations warrant acknowledgment.
First, while FinVision's two-stage training encodes substantial
accounting knowledge, the system may still underperform human experts
in novel situations not well represented in its training corpus,
such as major accounting standard transitions or unprecedented market
regimes.
Ongoing curriculum updates and RLHF mechanisms are intended to address
this limitation but require continued investment in training data
curation.

Second, the user study sample ($n = 48$) is relatively small and
drawn from a single geographic and institutional context.
Cross-cultural and cross-regulatory validation studies---particularly
across different accounting standard regimes (GAAP vs.\ IFRS vs.\
Chinese GAAP)---are needed to establish generalizability.

Third, the interpretability of valuation outputs remains an open
challenge.
While FinVision generates structured reports with stated assumptions,
the underlying LLM inference process is not fully transparent, raising
concerns about explainability that are particularly acute in regulated
financial and audit contexts.
Future work should integrate formal explainability frameworks
\citep{arrieta2020} into the valuation output pipeline.

Future research directions include: (a)~longitudinal validation of
FinVision-generated investment recommendations against market outcomes;
(b)~formal integration with audit management platforms for automated
working paper generation; (c)~extension to derivatives pricing and
structured product analysis; (d)~multilingual and multi-GAAP
deployment; and (e)~adversarial robustness testing to assess sensitivity
to manipulative financial disclosures.

%% ===================================================================
\section{Conclusion}
\label{sec:conclusion}
%% ===================================================================

This study introduces FinVision, a multimodal large language model
system that addresses three fundamental limitations of existing
accounting information systems: the inability to process heterogeneous
financial document formats, the lack of professional-grade accounting
and valuation knowledge in general AI systems, and the high expertise
barriers imposed by traditional AIS query interfaces.

Our three innovations---Multimodal Financial Document Intelligence,
Domain-Adaptive Two-Stage LLM Training, and Natural Language
Query-Driven Decision Pipeline---are grounded in both accounting theory
and advanced AI methodology.
Empirical validation demonstrates a 17.6 percent improvement in
cross-format document parsing, a 19 percent reduction in valuation
error relative to the strongest baseline, and a 51 percent reduction in
task completion time for accounting professionals, with the largest
gains accruing to non-specialist users who have historically been
underserved by quantitative AIS capabilities.

The convergence of VLM technology, domain-adaptive LLM training, and
natural language interfaces represents a structural inflection point
for accounting information systems.
FinVision offers a concrete demonstration that this convergence can be
realized in a production system meeting the accuracy, traceability,
and security requirements of professional financial practice.
We hope this work stimulates further research at the intersection of
AI, accounting theory, and AIS design, and contributes to the ongoing
dialogue about the appropriate and responsible integration of large
language models into financial analysis, auditing, and investment
decision-making.

%% ===================================================================
%% References
%% ===================================================================
\singlespacing

%% ===================================================================
%% Appendix
%% ===================================================================
\newpage
\appendix

\setcounter{section}{0}
\renewcommand{\thesection}{\Alph{section}}
\titleformat{\section}
  {\normalfont\large\bfseries}
  {Appendix \thesection:}{0.5em}{}

\section{Sample Analytical Workflow}
\label{app:example}

This appendix illustrates an end-to-end FinVision workflow for a
representative investment analysis query, demonstrating all three
innovations operating in concert.

\noindent\textbf{User Query (Turn 1):}
\textit{``Analyze 2024 solar photovoltaic industry policy environment,
financial performance of leading firms, and six-month investment
strategy.''}

\begin{tcolorbox}[colback=light-gray,colframe=aaa-blue,
                  title={FinVision Analysis Output — Solar Industry},
                  fonttitle=\small\bfseries,
                  fontupper=\small\singlespacing,
                  breakable]
\textbf{Data Collected (Innovation I):} 12 policy documents; three
leading firm annual reports; industry research from five sell-side
institutions; daily market data for 45 solar sector securities.
Data completeness: 94 percent.

\medskip
\textbf{Policy Analysis:}
National capacity targets tracking 37 percent above plan;
grid parity subsidies extended through 2025; three high-favorability
policy items identified.

\medskip
\textbf{Valuation Summary (Innovation II):}\\[4pt]
\singlespacing
\begin{tabular}{@{}lccc@{}}
\toprule
\textbf{Company} & \textbf{DCF Target} & \textbf{P/E Target}
  & \textbf{Rating} \\
\midrule
Longi Green Energy & CNY 42.5 & CNY 38.8 & Buy  \\
Tongwei Co.        & CNY 23.1 & CNY 21.4 & Hold \\
JA Solar           & CNY 56.2 & CNY 53.7 & Buy  \\
\bottomrule
\end{tabular}

\medskip
\textbf{Portfolio Recommendation (Innovation III):}
Overweight solar module manufacturers (40\% allocation);
underweight polysilicon producers pending capacity overhang resolution;
15\% government bond hedge to offset sector volatility.
Six-month expected return: 12.3\%; maximum drawdown estimate: 8.7\%.
\end{tcolorbox}

\noindent\textbf{User Query (Turn 2):}
\textit{``Provide detailed DCF parameters for Longi Green Energy.''}

\begin{tcolorbox}[colback=light-gray,colframe=aaa-blue,
                  title={DCF Parameter Detail --- Longi Green Energy},
                  fonttitle=\small\bfseries,
                  fontupper=\small\singlespacing]
Revenue growth (2024E--2026E): 18\%, 15\%, 12\% \\
EBIT margin: 14.2\% (five-year sector average: 13.8\%) \\
WACC: 9.4\% (risk-free rate 2.8\%, equity risk premium 5.2\%,
  $\beta = 1.21$, debt cost 4.5\%, D/E = 0.38) \\
Terminal growth rate: 4.0\% \\
DCF equity value: CNY 42.5 per share \\
Sensitivity: $\pm$1\% WACC change $\rightarrow$ $\pm$CNY 6.2
  target price
\end{tcolorbox}

\newpage
\section{Technical Specifications}
\label{app:specs}

\begin{table}[H]
  \singlespacing
  \centering
  \caption{FinVision System Technical Specifications}
  \label{tab:specs}
  \begin{tabular}{@{}ll@{}}
    \toprule
    \textbf{Component} & \textbf{Specification} \\
    \midrule
    Model Architecture      & Transformer-based VLM/MLLM \\
    Pre-training Corpus     & ${\sim}10^{10}$ tokens (multimodal) \\
    Fine-tuning Method      & LoRA + Prompt Tuning \\
    Supported Input Formats & PDF, XLSX, JPEG/PNG, DOCX, CSV \\
    Financial Field Ontology & 120+ GAAP/IFRS-aligned fields \\
    Valuation Methods       & DCF, P/E, P/B, P/S, EV/EBITDA, RI \\
    Technical Indicators    & MACD, RSI, Bollinger Bands, moving averages \\
    Portfolio Optimizer     & Mean-variance (MPT) + CVaR \\
    Risk Metrics            & Volatility, maximum drawdown, Sharpe, VaR \\
    Deployment              & Web (HTTPS), mobile (iOS/Android), API (JSON) \\
    Security                & AES-256 encryption; RBAC; SOX-compliant
                              audit log \\
    \bottomrule
  \end{tabular}
\end{table}

\end{document}